\documentclass[conference,a4paper]{IEEEtran} \IEEEoverridecommandlockouts
\IEEEaftertitletext{\vspace{-2\baselineskip}}%
\usepackage[font=footnotesize,belowskip=-2pt,aboveskip=4pt]{caption}
\usepackage{amsmath}
\usepackage[T1]{fontenc}
\usepackage{cite}
\usepackage{units}
\usepackage[dvips]{graphicx}
\usepackage{eepic}
\usepackage{overpic}
\usepackage[dvips]{color}
\usepackage{cite}
\usepackage{psfrag}
\newcommand{\titletext}{Tight Upper and Lower Bounds to the Information Rate of the Phase Noise Channel}

\begin{document}

\sloppy
\title{\titletext}

\author{
\authorblockN{Luca Barletta}
\authorblockA{Institute for Advanced Study\\
Technische Universit\"{a}t M\"{u}nchen \\
D-85748 Garching, Germany\\
Email: Luca.Barletta@tum.de} \and
\authorblockN{Maurizio Magarini, Arnaldo Spalvieri}
\authorblockA{Dipartimento di Elettronica, Informazione e Bioingegneria\\
Politecnico di Milano\\
I-20133 Milan, Italy\\
Email: \{Maurizio.Magarini,Arnaldo.Spalvieri\}@polimi.it}}

\maketitle
\begin{abstract}
Numerical upper and lower bounds to the information rate transferred through
the additive white Gaussian noise channel affected by
discrete-time multiplicative autoregressive moving-average (ARMA) phase noise are proposed in the
paper. The state space of the ARMA model being multidimensional,
the problem cannot be approached by the conventional trellis-based
methods that assume a first-order model for phase noise and
quantization of the phase space, because the number of state of
the trellis would be enormous. The proposed lower and upper bounds
are based on  particle filtering and Kalman filtering. Simulation
results show that the upper and lower bounds are so close to each
other that we can claim of having numerically computed the actual information
rate of the multiplicative ARMA phase noise channel, at least in
the cases studied in the paper. Moreover, the lower bound, which
is virtually capacity-achieving, is obtained by demodulation of
the incoming signal based on a Kalman filter aided by past data.
Thus we can claim of having found the virtually optimal
demodulator for the multiplicative phase noise channel, at least for the cases considered in the paper.
\end{abstract}

\section{Introduction}

Multiplicative phase noise is a major source of impairment in
radio and optical channels. The presence of phase noise in radio
channels is well known and studied from a long time, being phase
noise introduced by the local oscillators used in up conversion
and down conversion, while multiplicative phase noise is recently
becoming a hot topic in the context of coherent optical
transmission. Recent studies about the phase noise that arises in
optical channels and about its effects in coherent optics can be
found in \cite{essiambre,mag}. Several methods have been proposed in
the literature to combat the detrimental effects of phase noise.
Among these methods we cite iterative demodulation and decoding
techniques of \cite{shamai,colavolpe,barbieri} and the insertion
of pilot symbols \cite{spalv}, and staged demodulation
and decoding \cite{staged}.

The capacity of the additive white Gaussian noise (AWGN) channel
affected by multiplicative phase noise with white power spectral
density is studied in \cite{essiambre},\cite{goebel,hou}, while
Wiener's phase noise is considered in
\cite{barl,barl2,furboni,dauwels}. Analytical bounds on capacity of phase noise channels at high signal-to-noise ratio are given in \cite{lapidoth}. Despite the quantity and
quality of the literature available, we find room for new results
by considering the channel impaired by autoregressive moving-average (ARMA) multiplicative phase
noise, a phase noise model that is much more realistic than
Wiener's phase noise and/or white phase noise in many cases of
practical interest. The ARMA model makes it possible to shape the
power spectral density of phase noise by acting on the order and
on the parameters of the model. Working out the capacity of a
channel affected by a general multiplicative ARMA phase noise
process is a challenging problem, because
\begin{itemize}\item{the state space is not finite and it
is multidimensional, therefore it cannot be approached by
techniques like those used for white and Wiener phase noise,}
\item{the observation is a nonlinear function of the state.}
\end{itemize} The only paper studying the capacity of the
channel affected by ARMA phase noise we are aware of is
\cite{dauwels}, where the method of particle filtering (see
\cite{particle} for a tutorial on particle filtering) is adopted
to work out an approximation to the constrained channel capacity,
the constrained capacity being the information rate transferred
through the channel with a fixed source.
The new results presented in this paper are tight numerical upper and lower
bounds to the constrained capacity of the AWGN ARMA phase noise
channel.

\section{First-Order Markov Channels with Continuous State}

Let $u_{i}^{k}$ indicate the column vector $
 (u_k, u_{k-1}, \ldots, u_i)^T$,  $i\le k$,
where $u_i^k$ is empty for $i>k$, the superscript $^T$ denotes
transposition, and $u_i^k \in {\cal U}_i^k$. Also, let $U$
indicate a possibly non-stationary process, $U=(U_0,U_1, \cdots)$,
whose generic realization is the sequence $(u_0, u_1, \cdots).$
When ${\cal U}_i^k$ is a continuous set, $p(u_i^k)$ is used to
indicate the multivariate probability density function, while when
${\cal U}_i^k$ is a discrete set $p(u_i^k)$ indicates the
multivariate mass probability and $|{\cal U}_i|$ denotes the
number of elements in ${\cal U}_i$.

Consider a first-order Markov channel. The Markovian state process
$S$ is characterized by the joint probability
\begin{equation}\label{markovstate}
p(s_0^{n}) = p(s_{0})\prod_{k=1}^n p(s_k|s_{k-1}).
\end{equation}
A channel without feedback that is memoryless given the state is characterized by
the state transition probability $p(s_k|s_{k-1})$ and by the
conditional distribution
\begin{align}
p(y_1^n|x_1^n,s_1^n)&=\prod_{k=1}^np(y_k|x_k,s_k) ,
\label{markovchannel}
\end{align} where $Y$ is the channel output process and $X$ is the channel input process, that we assume to be discrete.
Equation (\ref{markovchannel}) says that the channel output
process is memoryless given the source and the state. Drawing from
the parlance of carrier recovery, the channel transition
probability $p(y_k|x_k,s_k)$, which is conditioned on channel's
input, is hereafter called {\em data-aided} channel transition
probability. We assume that the source is memoryless and
independent of the state, that is
\begin{align}
p(x_1^n|s_1^n)&=\prod_{k=1}^np(x_k). \label{markovsource}
\end{align}
Putting together (\ref{markovchannel}) and (\ref{markovsource})
one finds that the joint source and channel model is memoryless
given the state:
\begin{align}
p(y_1^n,x_1^n|s_1^n)&=\prod_{k=1}^np(y_k,x_k|s_k).
\label{markovjoint}
\end{align}
Using (\ref{markovjoint}) one finds that channel's output is
memoryless given the state:
\begin{align}
p(y_1^n|s_1^n) &=\sum_{x_1^n \in {\cal X}_1^n}
p(y_1^n,x_1^n|s_1^n) =
 \sum_{x_1^n \in {\cal
X}_1^n}
\prod_{k=1}^np(y_k,x_k|s_k) \nonumber \\
& = \prod_{k=1}^n \sum_{x_k \in {\cal X}_k} p(y_k,x_k|s_k) =
\prod_{k=1}^{n} p(y_k|s_k). \label{memorylesschannel}
\end{align}
Drawing again from the parlance of carrier recovery, the channel
transition probability $p(y_k|s_k)$, which is not aware of
channel's input, is hereafter called {\em blind} channel
transition probability. From eq. (\ref{markovstate}) and
(\ref{memorylesschannel}), after straightforward passages one gets
\begin{equation}\label{markov}
p(s_k|s_{k-1},y_1^{k-1}) = p(s_k|s_{k-1}), \ \ k=1,2, \cdots, n.
\end{equation}
Also, by (\ref{markovjoint}) and (\ref{memorylesschannel}) one
finds that the source is memoryless given the state and channel's
output:
\begin{align}
p(x_1^n|y_1^n,s_1^n) = \prod_{k=1}^{n} p(x_k|y_k,s_k).
\label{memorylesssource}
\end{align}

\section{Bayesian Tracking}

Any measurement process $Y$ that is memoryless given the state can
be cast in the general framework of state-space approach for
modelling dynamic systems, which is defined by the state
transition equation
\begin{equation}s_{k}=f_{k}(s_{k-1},v_{k-1}), \ \ k=1,2, \cdots, n,\label{statetran}
\end{equation} and by the measurement equation
\begin{equation}
y_k=h_k(s_k,n_k), \ \ k=1,2, \cdots, n,\label{measurement}
\end{equation} where $f_k(\cdot)$ and $h_k(\cdot)$ are possibly
non-linear and time-varying known functions of their arguments,
$v_k$ is the process noise vector, and $n_k$ is the measurement
noise vector, which is assumed to be independent of $v_k$. The
state-space approach fits the Markov channel, taking the output
channel process $Y$ as the measurement process both in the blind
and in the data aided case. In the blind case, the measurement
equation is a time-invariant function of the state, and the
measurement noise is the joint effect of channel noise and input
process. The blind case is described by the memoryless probability
$p(y_k|s_k)$ appearing in the product (\ref{memorylesschannel}).
In the data-aided case the measurement noise is only the channel
noise and the input process is embedded in the known non-linear
and time-varying $h_k(\cdot)$. In this case the measurement
probability is $p(y_k|x_k,s_k)$.

A powerful tool in the analysis of dynamical system is the
so-called {\em Bayesian tracking}. Let the Markovian state be
continuous. One can track the hidden state by a two-step recursion
that, for $k=1,2,\cdots,n,$ reads
\begin{equation}
p(s_k|y_1^{k-1}) = \int_{{\cal
S}}p(s_k|s_{k-1})p(s_{k-1}|y_1^{k-1}) d s_{k-1}, \label{predict}
\end{equation}
\begin{equation} p(s_k|y_1^{k})
=\frac{p(s_k|y_1^{k-1})p(y_k|s_k)}{p(y_k|y_1^{k-1})}, \ \ \
\label{update}
\end{equation}
where $p(s_k|y_1^{k-1})$ is the {\em predictive} distribution,
$p(s_k|y_1^k)$ is the {\em posterior} distribution, and the
denominator of (\ref{update}) is a normalization factor such that
the left-hand side is a probability. The normalization factor can
be computed by the Chapman-Kolmogoroff equation
\begin{equation}
p(y_k|y_1^{k-1}) =\int_{{\cal S}_k} p(s_k|y_1^{k-1})p(y_k|s_k) d
s_{k}. \label{ck}
\end{equation}
The state transition  probability $p(s_k|s_{k-1})$ appears in
(\ref{predict}) in place of $p(s_k|s_{k-1},y_1^{k-1})$ thanks to
(\ref{markov}). Thanks to (\ref{memorylesschannel}), $p(y_k|s_k)$
can be used in place of $p(y_k|s_k,y_1^{k-1})$ in (\ref{update}).

When the dynamic system is a linear system with Gaussian noises,
Bayesian tracking is performed by the Kalman filter. When the
model is not tractable, one can resort to particle filtering
techniques to work out an approximation to the wanted
distribution.

The probabilities worked out by Bayesian tracking can be used to
evaluate entropy rates by Monte Carlo integration as, for instance, in
\cite{dauwels, barl2}. When the result of Bayesian tracking is an
approximation $q(u_1^k)$  to the wanted probability $p(u_1^k)$, then, by the
Kullback-Leibler inequality, the approximation can be used to get
an upper bound on the wanted entropy rate
\begin{equation}\overline{h}(U)= -\lim_{k \rightarrow \infty}
\frac{1}{k}E_p \left\{\log_2 q(u_1^{k})\right\} \geq
h(U),\label{klbound} \end{equation}
where operator $E_p$ denotes expectation with respect to probability $p(\cdot)$.

\section{The ARMA Phase Noise Channel}

The $k$-th output of the channel is
\begin{equation} y_k=x_ke^{j \phi_k}+w_k, \ \ \ k=1,2,\cdots,n\label{envelope}
\end{equation} where $j$ is the imaginary unit,
$Y$ is the complex channel output process, $X$ is the channel
complex input modulation process made by i.i.d. random variables
with zero mean and unit variance, $W$ is the complex AWGN process
with zero mean and variance $\mbox{SNR}^{-1}$, and $\Phi$ is the
phase noise process which is assumed to be independent of $X$ and
$W$. Specifically, process $\Phi$ is modelled as the 1-causal
accumulation modulo $2 \pi$ of frequency noise, that is
\begin{equation}
\phi_{k}= [\lambda_{k-1}+\phi_{k-1}]_{\bmod{2\pi}}, \label{nco}
\end{equation}
where the frequency noise process $\Lambda$ is given by the $z$-transform
\begin{align}\sum_{k=-\infty}^{\infty}\lambda_kz^{-k}=H(z)
\cdot \left( \sum_{k=-\infty}^{\infty}v_kz^{-k} \right)\nonumber
\end{align} where $V$ is a white Gaussian noise process with zero mean
and variance $\gamma^2$, and
\begin{align}H(z)=\frac{\prod_{k=1}^N(1-\beta_{k}
z^{-1})}{\prod_{k=1}^N( 1-\alpha_{k} z^{-1})}=\frac{1+\sum_{k=1}^N
b_{k} z^{-k}}{ 1-\sum_{k=1}^N a_{k} z^{-k}}, \label{innovation}
\end{align}
where $|\alpha_{k}|<1, \ |\beta_{k}| \leq 1, \ N \geq 0$, and it is understood that $H(z)=1$ for $N=0$, leading to the
special case of random phase walk, where $\lambda_k=v_k$. $H(z)$ is the transfer function of a filter made by a shift
register with feedback taps $a_1^N$ and forward taps $b_1^N$.  Let
$\omega_{k-N}^{k-1}$ be the content of the shift register at the
$k$-th channel use, that is
\[\sum_{k=-\infty}^{\infty}\omega_kz^{-k}=
\frac{\sum_{k=-\infty}^{\infty}v_kz^{-k}}{1-\sum_{k=1}^N a_{k}
z^{-k}}.\] The state at time $k$ is the $(N+1)$ column vector
\begin{equation}
s_{k}=(\phi_k,(\omega_{k-N}^{k-1})^T)^T. \label{state}
\end{equation}
 Let us introduce the state transition matrix
\begin{align}
      F= \left[\begin{array}{ccc}
         1 & \multicolumn{2}{c}{(a_1^N+b_1^N)^T} \\
         0 & \multicolumn{2}{c}{(a_1^N)^T}  \\
         0_{N-1} & I_{N-1} & 0_{N-1}
        \end{array}\right] \nonumber,
\end{align}
where $I_N$ is the identity matrix of size $N \times N$ and $0_N$
is a column vector of $N$ zeros. The state transition equation is
\[s_{k+1}=Fs_k+(v_k,v_k,0_{N-1}^T)^T +(2 m \pi,0_{N}^T)^T,\]
where $m$ is such that $\phi_{k+1}$ lies in the interval
$[0,2\pi)$, thus making the state transition equation non-linear.

Given $s_{k}$, for $N=0$ the state transition to $s_{k+1}$ is
ambiguous of $2 n \pi$, while for $N \geq 1$, due to the presence
of $\omega_{k}$ in $s_{k+1}$, the state transition is not
ambiguous. Although not necessary, in the following we will assume
$N\geq 1$, referring the reader to \cite{barl2} for the state
transition probability with $N=0$. For $N\geq 1$ the state
transition probability is a $(N+1)$-dimensional Gaussian
distribution. Note that, given $s_k$, $N$ of the $(N+1)$ entries
of $s_{k+1}$ are known, the only free random variable being $v_k$,
hence the covariance matrix of the state transition probability
has unit rank. Specifically,
\begin{align} \label{statetransition}
 p(s_{k+1}|s_k) =  g_{N+1}
 (Fs_k+(2m\pi,0_N^T)^T,\Sigma_v; s_{k+1}),
\end{align}
where $g_N(\mu, \Sigma; x)$ is a $N$-dimensional Gaussian
distribution over the space spanned by $x$ with mean vector $\mu$
and covariance matrix $\Sigma$,
\begin{align}
      \Sigma_v=  \left[\begin{array}{ccc}
         \gamma^2  & \gamma^2  & 0_{N-1}^T  \\
         \gamma^2  & \gamma^2  & 0_{N-1}^T  \\
         0_{N-1} & 0_{N-1} & 0_{(N-1)\times(N-1)}
        \end{array}\right] \label{qmatrix},
\end{align}
where $0_{N \times M}$ is an all-zero $N \times M$ matrix, and
\begin{equation}\label{moduloconstraint}
2 m \pi= \phi_{k+1}- \phi_k - \omega_k - \sum_{i=1}^N b_i
\omega_{k-i}.
 \end{equation}

The measurement at time $k$ is the $y_k$ given by
(\ref{envelope}). The data-aided channel transition probability is
\begin{equation}p(y_k|x_k,s_k)= g_c(x_ke^{j \phi_k}, \mbox{SNR}^{-1};y_k),
\label{channelprob}
\end{equation}
where $g_c(\mu,\sigma^2;t)$ indicates a circular symmetric Gaussian probability
density function over the complex plane spanned by $t$ with mean
$\mu$ and two-dimensional variance $\sigma^2$. The joint source
and channel probability is
\begin{equation}
p(y_k,x_k|s_k)=p(x_k) g_c(x_ke^{j \phi_k}, \mbox{SNR}^{-1};y_k).
\label{sourcechanneltransition}
\end{equation}
From the above probability one can compute the  blind channel
transition probability by (\ref{memorylesschannel}).

\section{Upper Bound}

Let $h(U)$ denote the entropy rate of process $U$. Extract $h(Y|X)$ from \[ h(Y|X)-h(Y|X,S)=h(S|X)-h(S|X,Y),
\]
to write
\begin{equation} I(X;Y)=h(Y)-h(Y|X,S)-h(S)+h(S|X,Y),\label{i1}
\end{equation} where, by independence between $X$ and the state process $S$, $h(S)$
has been substituted in place of $h(S|X)$.
 The upper bound
that we propose is
\begin{equation} \label{full}
\overline{h}(Y)-h(Y|X,S)-h(S)+ \overline{h}(S|X,Y) \geq I(X;Y),
\end{equation}
where $\overline{A}$ indicates an upper bound on $A$.
The two relative entropy rates $h(Y|X,S)$ and $h(S)$ are those of
the white Gaussian processes $W$ and $V$, respectively. The upper
bound $\overline{h}(Y) \geq h(Y)$ can be obtained by approximating
the conditional probability $p(y_k|y_1^{k-1})$ to the
normalization factor of blind Bayesian tracking performed by a
particle filter as in \cite{dauwels}.

The new contribution of the present paper is the upper bound
$\overline{h}(S|X,Y)$, which is worked out as follows. Invoking
the chain rule, the Markovian property (\ref{markov}), and the
Shannon-McMillan-Breiman theorem, one can evaluate the entropy
rate by computer simulation as
\begin{equation}
h(S|X,Y)= \lim_{n\rightarrow\infty} \frac{1}{n}\sum_{k=1}^n -
\log_2 p(s_k|x_1^k,y_1^k,s_{k+1}), \label{SMB2} \end{equation}
where $(x_1^n,y_1^n,s_1^{n+1})$ is a realization of the joint
process $(X,Y,S)$. Unfortunately, the actual
$p(s_{k}|x_{1}^{k},y_1^k,s_{k+1})$ of (\ref{SMB2}) is not tractable. We propose to
approximate it as
\begin{equation}q(s_{k}|x_{1}^{k},y_1^k,s_{k+1})=\frac{p(s_{k+1}|s_{k})q(s_{k}|x_1^{k},y_1^k)
}{ \int_{\cal S}p(s_{k+1}|s_{k})q(s_{k}|x_1^{k},y_1^k) d
s_{k}},\label{app} \end{equation} with
\begin{align}
q(s_k| x_1^k,y_1^k)  = \sum_{l=-\infty}^{\infty} g_{N+1}(\mu_k,
\Sigma_k; \phi_k+
 2 l \pi,
\omega_{k-N}^{k-1}), \label{postkalman}
\end{align}
thus, thanks to (\ref{klbound}), getting the upper bound
\begin{equation}
\overline{h}(S|X,Y)= \lim_{n\rightarrow\infty}
\frac{1}{n}\sum_{k=1}^n - \log_2 q(s_k|x_1^k,y_1^k,s_{k+1}).
\nonumber
\end{equation}
 The denominator of (\ref{app}) can be
treated by moving the sum (\ref{postkalman}) outside the integral,
and observing that the integral is the convolution between two
Gaussian distributions, leading to closed form computation as in
the predictive step of the Kalman filter
\cite[Sec. 3.3]{simon}:
\begin{align} & \int_{\cal
S}p(s_{k+1}|s_{k})q(s_{k}|x_1^{k},y_1^k)\ d
s_{k}=  \nonumber \\
&\hspace{-0.15cm} \sum_{l=-\infty}^{\infty} g_{N+1}(F \mu_{k},F\Sigma_{k}F^T
\hspace{-0.15cm}+ \Sigma_v; \phi_{k+1}+
 2 l \pi,
\omega_{k-N+1}^{k}).\label{predictcovkalman}
\end{align}
The parameters $\mu_k$ and $\Sigma_k$ appearing in equations
(\ref{postkalman}) and (\ref{predictcovkalman}) can be worked out
by a linearized Kalman filter \cite[Sec. 13.2]{simon}. As it will be shown by simulation
results, a tighter bound can be obtained by taking for $\mu_k$ and
$\Sigma_k$ a sample estimate where the sample is the set of
posterior particles of a particle filter. Note that the integral
in the denominator of (\ref{app}) is a normalization factor such
that the left side of (\ref{app}) is a probability. As a
consequence, it cannot be evaluated by the predictive particles of
the particle filter, because the predictive particles would
provide only an approximation to the wanted integral, and using an
approximation to the denominator is not sufficient to guarantee
that the ratio in (\ref{app}) is a probability. Also, it is worth
pointing out that, while in \cite{dauwels} the phase in the state
model is unwrapped, here it is the evaluation of
$\overline{h}(S|X,Y)$, that is not made in \cite{dauwels}, that
forces us to define the state by the wrapped phase (\ref{nco}). As
a matter of fact, phase ambiguities of $2n\pi$ are inherently
present in the measurement, therefore cycle slips of the Bayesian
tracking algorithm would lead to catastrophic errors of $2 n \pi$
between the actual unwrapped phase and the distribution of the
unwrapped phase recovered by the tracking algorithm.

\section{Lower Bound}
Assume a discrete input alphabet. The lower bound that we propose is
$H(X)-\overline{H}(X|Y) \leq I(X,Y)$, where, by  the same arguments leading to (\ref{SMB2}) and by the
Kullback-Leibler inequality (\ref{klbound}), one evaluates the upper bound on the
conditional entropy rate as
\begin{align} \overline{H}(X|Y) = \lim_{n \rightarrow \infty}
\frac{1}{n} \sum_{k=1}^n - \log_2 q(x_k|x_{1}^{k-1},y_1^{n}).
\label{lb}
\end{align}
The upper bound can be based on demodulation, that is on the
probability
\begin{align} \label{saturation}p(x_{k}| x_1^{k-1}, y_1^{n})=
\int_{ {\cal S}} p(s_k,x_{k}|x_1^{k-1}, y_1^n) ds_k,
\end{align}
where the probability inside the integral can be written as
\begin{align} \label{twoterms}p(s_k, x_{k}|x_1^{k-1},  y_1^n)
&= p(s_k| x_1^{k-1}, y_1^n)p(x_{k}|s_k,x_1^{k-1}, y_1^n) \nonumber
\\
& =p(s_k|x_1^{k-1},y_1^n)p(x_{k}|s_k, y_k),
\end{align}
where the second equality comes from (\ref{memorylesssource}). In
what follows the first factor in (\ref{twoterms}) is approximated
to $p(s_k| x_1^{k-1}, y_1^{k})$. We point out that the proposed
approximation is likely to be tight, because the condition
$y_{k+1}^n$ gives only a weak contribution of non-data-aided type
to the wanted probability. The proposed approximation leads to
\begin{align} \label{saturation2} %\hspace{-0.1cm}
& q(x_{k}| x_1^{k-1}, y_1^{n})= \int_{ {\cal S}} q(s_k|x_1^{k-1},
y_1^{k})p(x_k|s_k,y_k) ds_k \nonumber \\ 
% & \hspace{-0.1cm}=\int_{
% {\cal S}} \frac{q(s_k|x_1^{k-1},
% y_1^{k-1})p(y_k|s_k,x_1^{k-1},y_1^{k-1})}
% {p(y_k|x_1^{k-1},y_1^{k-1})}p(x_k|s_k,y_k) ds_k \nonumber \\
& %\hspace{-0.1cm}
=\int_{ {\cal S}} \frac{q(s_k|x_1^{k-1},
y_1^{k-1})p(y_k|s_k)}
{p(y_k|x_1^{k-1},y_1^{k-1})}\frac{p(y_k,x_k|s_k)}{p(y_k|s_k)}
ds_k \nonumber \\
& %\hspace{-0.1cm} 
\propto \int_{ {\cal S}} q(s_k|x_1^{k-1},
y_1^{k-1})p(y_k,x_k|s_k) ds_k,
\end{align}
which, after normalization, can be used in (\ref{lb}) to get the
desired bound. The first factor inside the integral
(\ref{saturation2}) is the predictive probability of Bayesian
tracking, while the second factor is a memoryless term that comes
from the channel model (\ref{sourcechanneltransition}).

\section{Simulation Results}

\begin{figure}[!thbp]
   \centering
   \includegraphics[width=.95\columnwidth]
   {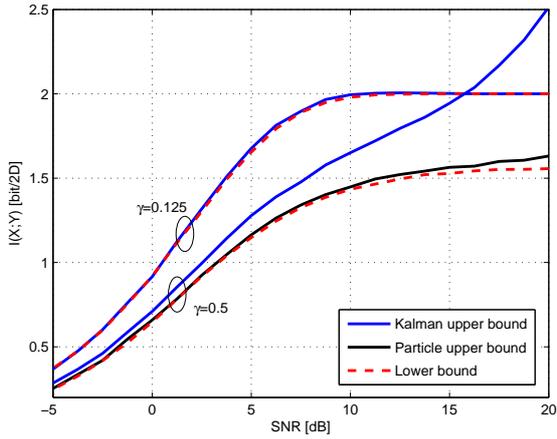}
   \caption{Upper and lower bounds to the information rate of
   4-QAM.
   }   \label{4} \vspace{-.4cm}
\end{figure}

\begin{figure}[!thbp]
   \centering
   \includegraphics[width=.95\columnwidth]
   {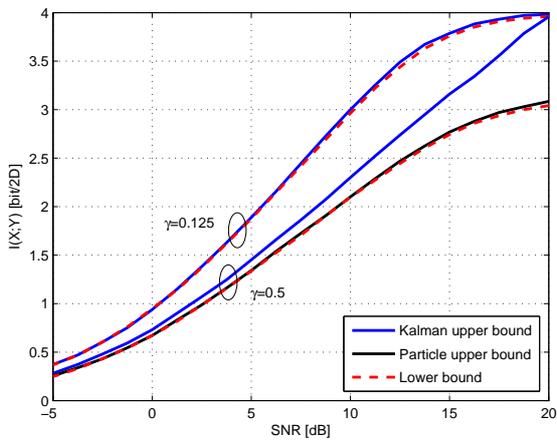}
   \caption{Upper and lower bounds to the information rate of
   16-QAM.   }   \label{16}
\end{figure}

The frequency noise used in the simulations is obtained by
filtering white Gaussian noise through the transfer function
\begin{equation}
H(z)=\frac{(1-\beta_1 z^{-1})(1-\beta_2z^{-2})}{1-\alpha_1
z^{-1}}. \label{smmodel}
\end{equation}
 Special
cases of (\ref{smmodel}) are obtained with $\beta_1=\alpha_1$ and
$\beta_2=1$, leading to white phase noise, and $\beta_1=0, \
\beta_2=1, \ \alpha_1=1$, that leads to Wiener's phase noise. Model (\ref{smmodel}) is proposed in \cite{SM} as an approximation to the phase
noise spectrum of real-world microwave local oscillators and it
has been used  with $\alpha_1 = 0.9999$, $\beta_1=0.9937$,
$\beta_2=0.7286$ to get the simulation results that are hereafter
presented. The lower bound is computed by adopting as a Bayesian
tracking method the linearized predictive Kalman filter, as in
\cite{oekalman} and \cite{indo}, while for the upper bound we use
both the Kalman filter and the particle filter. Figure \ref{4}
reports the results for 4-ary quadrature-amplitude modulation (QAM) while Fig. \ref{16} reports the
results for 16-QAM, in both cases with two values of $\gamma$. The
two Figures show that the particle filter greatly improves the
upper bound over the Kalman filter, especially for large $\gamma$.
In contrast, the lower bound based on the predictive Kalman filter
is so tight that there is no need of using a particle filter for
demodulation, also for large values of $\gamma$. We have observed
that the Kalman filter often produces a covariance $\Sigma_k$ with
a determinant that is much lower than the one that is obtained by
the particle filter. What happens is that the folded Gaussian
distribution (\ref{postkalman}) is sampled in the state visited by
the simulation, and, when this state is far from the mean vector,
the Gaussian is sampled on the tails. In this event, the poor
estimation of the covariance leads to dramatically large errors in
the evaluation of the differential entropy rate $h(S|X,Y)$.
Conversely, the entropy rate $\overline{H}(X|Y)$ that appears in
the lower bound is based on the integral of the mentioned Gaussian
distribution, hence it is less sensitive to errors in the
estimated covariance.
\vspace{-5mm}

\section{Conclusion}
We have presented upper and lower bounds to the constrained
information rate transferred through the multiplicative phase
noise channel with ARMA phase noise. From the results it appears
that the upper and lower bounds are so close to each other that we
can claim of having computed the actual information rate, at least
for the second-order ARMA phase noise studied in the simulation.
An important experimental result presented in the paper is that
demodulation based on a predictive linearized Kalman filter aided
by past data is virtually capacity achieving, at least in the
examples studied in the paper. This is not surprising in view of
the result obtained in \cite{forney} for the intersymbol interference (ISI) channel, that
says that predictive filtering aided by past data (in the case of
the ISI channel, the predictive decision-feedback equalizer) virtually leads to channel
capacity. A practical mean to replace past data with the decisions
coming from a capacity achieving code is the interleaving scheme
originally proposed by Eyuboglu in \cite{eyuboglu} for the ISI
channel. Extension of this principle to other channels can be
found, for instance, in \cite{collins}. Computational complexity of demodulation via Kalman filter can be lowered by using a time invariant filter as described in \cite{bridging}.

\end{document}